\documentclass[10pt]{article}

\usepackage{amsfonts}

\def\c{\mathcal}

\def\n{\nabla}
\def\pa{\partial}

\def\be{\begin{equation}}
\def\ee{\end{equation}}
\def\ba{\begin{eqnarray}}
\def\ea{\end{eqnarray}}
\def\ba*{\begin{eqnarray*}}
\def\ea*{\end{eqnarray*}}
\def\bc{\begin{center}}
\def\ec{\end{center}}
\def\R{I \kern-.36em R}
\def\E{I \kern-.36em E}
\def\F{I \kern-.36em F}

\begin{document}

\title	{REMARKS ON THE ENTROPY OF NON-STATIONARY BLACK HOLES}
\author{G. Allemandi\footnote{E-mail: allemandi@dm.unito.it}, L. Fatibene\footnote{E-mail:
fatibene@dm.unito.it}, M. Francaviglia\footnote{E-mail: francaviglia@dm.unito.it}\\ Dipartimento di
Matematica, Universit\`a degli Studi di Torino, \\
Via Carlo Alberto 10, 10123 Torino, Italy}
\maketitle

\abstract{
The definition of entropy obtained for stationary black holes  is extended in this paper
to the case of non-stationary black holes. Entropy 
is defined as a macroscopical thermodynamical  quantity which satisfies the first 
principle of 
thermodynamics. In the non-stationary case a volume term appears since the solution 
does not admit a Killing vector.}

\section{\large Introduction}
It is known that in many situations the entropy of a black hole solution of Einstein's equations  can be calculated 
in the framework of a  classical field theory without resorting to a 
statistical approach.\\
This may be considered a good feature at present since statistical approaches are based on a 
Hamiltonian and/or quantum formulation and both these aspects are not yet clear 
for General Relativity.
The definition of black hole entropy for stationary black holes as a 
macroscopical quantity which satisfies the first principle of thermodynamics 
has in fact been proposed in \cite{IW1}, it has been settled on a secure mathematical ground in
  \cite{FF2} and supported by a lot of 
examples \cite{FF9},\cite{FF10}. In this paper we aim to discuss a 
possible extension to non-stationary cases.\\
\noindent We use the framework of classical field theory in its Lagrangian 
formulation; we require 
the theory to be natural, which means that each diffeomorphism of 
spacetime 
is a symmetry for the Lagrangian. For our purpose it is useful to rely on the 
geometric 
language of fiber bundles in which the calculus of variations is most naturally 
defined. In this framework, because of the first principle of 
thermodynamics, the 
product between the "temperature of the black hole horizon" and the "variation of 
entropy" is equal to an integral of a suitable $(n-2)$-form at space infinity 
(in a space time of dimension $n \geq 3$). 
The integrated form
$\alpha(L,\sigma,\xi,X)$ is obtained as the variation of a suitable conserved quantity
 (in the sense of N\"other's theorem) and it is
associated  to 
a vector $\xi=\pa_t+\Omega \pa_\phi$ on spacetime. 
In the case of stationary black holes, it is possible to transform the 
integral 
at space infinity into an integral on a trapping surface $\Pi$ for the singularities, 
since the 
form 
$\alpha$ turns out to be closed:
\be
T \delta_X S= \int_\Pi \alpha(L,\sigma,\xi,X) \label{prpr}
\ee
If the horizon of the black hole is bifurcate and we choose $\Pi$ to coincide with the 
bifurcation surface, we obtain, as a particular case, the same formula obtained by 
R. Wald and V. Iyer in \cite{IW1}. We stress however that the above assumptions on 
$\Pi$ are unnecessary, often difficult to deal with and sometimes impossible 
(see ref.s \cite{FF2}, \cite{FF9}, \cite{FF10}).

In this paper we try to extend the same definition of entropy  to the 
case of 
non-stationary black holes. We here consider  black holes   with an 
oscillating horizon without a 
quadrupole 
momentum, so that they do not emit gravitational waves 
(see, for example, R. Wald and V. Iyer \cite{IW1} or Frolov 
\cite{FR1},\cite{FR2}), for which the first principle of black holes thermodynamics has the 
form:
\be
\delta_X M = T \delta_X \c{S} + \Omega \delta_X {J} 
\ee 
because they can be considered as isolated systems. \\
The problems which occur in the non-stationary case are related to the fact 
that 
the vector $\xi$ is no longer a Killing vector for the solution $\sigma$, a 
fact that in stationary case is extensively used to prove that the form $\alpha$ is closed 
(see ref.s \cite{IW1}, \cite{FF2}). When we try to transform the integral 
at spatial infinity into an integral on a "finite surface", a volume 
integral 
of the so-called \textit{symplectic form} $\omega(\sigma,\xi,X)$ appears because 
the form $\alpha$ is no longer closed, i.e. ${Div} \> \alpha (L,\sigma,\xi,X) 
\ne 
0$. In this case the entropy is defined as the sum of two integrals:

\be
T \delta_X S_{dyn}= \int_\Pi \alpha(L,\sigma,\xi,X)+\int_\Sigma 
\omega(\sigma,\xi,X)
\ee
\noindent
where $\Sigma$ is the volume of the region enclosed between  $\Pi$ and space infinity. 
We will show that both integrals are well defined and can be easily 
evaluated in the framework we use.\\
An earlier proposal for dynamic black hole entropy was given by R. Wald and V. Iyer 
\cite{IW1}. 
They just tried to adapt the definition given for the stationary case but, in 
the end, their definition turned out not to be covariant due to 
unessential requirements made on the existence and structure of the 
horizons (see \cite{FF2}). The new definition we propose satisfies all the conditions imposed 
by R. Wald and V. Iyer in \cite{IW1} and in addition, as a consequence of 
the geometrical framework we use, our proposal is automatically covariant with 
respect to any fibered morphism, i.e. any redefinition of fields. This is a 
stringent requirement from a physical viewpoint, since any lack of 
covariance produces results which are either wrong or at least require a 
lot of efforts in order to get a correct physical interpretation.\\
A further problem, which we are not able to overcome, is due to the lack of examples 
on which one can test any prescription for entropy of non-stationary black holes. Basically we don't 
know any non-stationary and geometrically well defined exact solution, for which one has a reasonable physical interpretation. 
For this reason we shall not present any example and direct application of our 
framework. Nevertheless we believe that the result we obtain is of interest, since it enlights the 
concept of entropy even in the stationary case by clariffing which are the fundamental properties of entropy 
and which are instead mere consequences of stationarity.\\
Furthermore, even if we do not know any explicit solution to test the formalism, 
we stress that the class of solutions that are under consideration is certainly not empty, physically relevant 
and, as mentioned above, it has been taken into account in the literature 
(see references \cite{IW1}, \cite{FR1}, \cite{JP1}).

\section{\large Notation and review of the stationary case}
Hereafter we recall briefly the standard notation and the definition of 
entropy in the the stationary case (more details can be found in 
\cite{FF2}, \cite{FF1}, \cite{GG1}, \cite{GG2}). Let us consider a configuration  bundle $B$ 
fibered on a spacetime $M$ and let us denote by $J^k (B)$ its $k$-order 
jet bundle, i.e. the space where fields live together with their 
derivatives up to order $k$ included. Fibered local coordinates on $J^k (B)$ are defined by 
$(x^\mu, y^i,{y^i}_\mu, ...,{y^i}_{\mu_1,...,\mu_k})$. Let us also denote 
by $\Lambda^n (T^* M)$ the bundle of $n$-forms over $M$. A \textit{Lagrangian 
of order} $k$ defined on $B$ is a morphism of fiber bundles:
\be
L:J^k (B) \to \Lambda^n (T^* M) \label{qua}
\ee
In the case of General Relativity in vacuum we can choose the second order
 Hilbert Lagrangian:
\be
L=\frac{1}{2 k} R \sqrt{g} d s \label{H}
\ee
The variation of the generic Lagrangian (\ref{qua}) can be expressed through the so called 
\textit{first variation 
formula}. We consider a vertical vector $X$ on  $B$ and the variation of 
$L$ 
along the flow of $X$, evaluated on a section $\sigma$ of $B$:
\ba*
<\delta L \circ j^k \sigma| j^k X >= 
< \E (L) \circ j^{2k} \sigma| X > + d [ < \F ( L, \gamma) 
\circ j^{2k-1} \sigma | j^{k-1} X >] \label{varpr}
\ea*
where $\E (L)$ and $\F ( L, \gamma)$ are well-defined global morphisms. The Euler-Lagrange 
morphism:
\be
\E (L): J^{2k}(B) \to \Lambda^n (M) \otimes V^* (B) 
\ee
is unique and it defines the field equations $ \E (L) \circ   j^{2k} 
\sigma=0$. In this case $V^* (B)$ denotes the dual bundle of the vector 
bundle $V (B)$ of vertical vectors on $B$. The Poincar\'e-Cartan morphism $\F (L,\gamma)$ depends in general on the 
Lagrangian 
and on an arbitrary background connection $\gamma$ on $M$:
\be
\F (L, \gamma): J^{2k-1} (B) \to \Lambda^{n-1} (M) \otimes V^* (J^{k-1} B)
\ee
 \\
In particular, the Euler-Lagrange morphism for General Relativity gives 
the vacuum Einstein equation
\be 
<\E(L)   |  X>= 
  e_{\mu \nu}  
 \delta  g^{\mu  \nu }  d s= (R_{\mu \nu}  - 
\frac{1}{2} \sqrt{g} g_{\mu \nu}) \delta  g^{\mu  \nu }  d s \label{E1}
\ee
while the expression of the  Poincar\'e-Cartan morphism is in this case
\be 
<\F(L)   | j^{1} X>= 
P^{\rho \theta }_{\alpha \beta}  \> \nabla_\theta
\delta  g^{\beta  \alpha } \> d s_\rho \label{S2}
\ee
where we have set
\be
P^{\rho \theta }_{\alpha \beta}=
- \Big( \frac{1}{16 \pi G} \Big) \sqrt{g}  \;
[ g^{\rho \theta }  g_{\alpha \beta} -
 \delta^\rho_{(\alpha}   \delta^\theta_{\beta) } ] 
\ee
We can notice the Poincar\'e-Cartan morphism (\ref{S2}) does not depend on 
any background connection due to the low order ($k=2$) of the theory.\\
We recall also that one can introduce a definition for the Lie derivative of bundle sections 
with 
respect to the flow of a vector $\Xi$ on $B$ projectable over $\xi$ on $M$ as:
\be 
\pounds_\Xi \sigma=T \sigma (\xi)- \Xi \circ \sigma \label{lie}
\ee
A projectable vector field $\Xi$ is an infinitesimal 
symmetry of the Lagrangian iff the following holds:
\be
< \delta L \circ j^k \sigma \mid j^k \pounds_\Xi \sigma > = 
{d} ( i_\xi   L)  \label{pot}
\ee
\noindent A bundle is \textit{natural} iff for each spacetime diffeomorphism 
$f:M \to M $ on the basis $M$ it is possible to find a canonical 
lift $\phi_f : B \to B$ on the bundle. \\
A field theory is natural iff the configuration bundle is natural and each  
diffeomorphism on  the basis $M$ is a symmetry of the Lagrangian $L$ 
(in the sense that $L$ is invariant under the pull back via any lift of $\phi_f$). In 
natural theories we can define a Lie derivative with respect to a spacetime 
vector field $\xi$ by setting:
\be 
\pounds_\xi \sigma= \pounds_\Xi \sigma
\ee
In this letter we will treat only natural theories, but the formalism 
introduced here is also valid for the more general case of gauge-natural theories \cite{FF2}, 
\cite{LF1}.\\

In the case of the Hilbert Lagrangian (\ref{H}) which gives to General 
Relativity the structure of a natural theory, the covariance 
condition (\ref{pot}) with respect to the vector $\xi$ can be expressed in the form:
\be
d_\rho (\xi^\rho L)= \frac{1}{2 } e_{\mu \nu} \pounds_\xi g^{\mu \nu}-  
\frac{1}{2 k}  \sqrt{g} g^{\alpha \beta} \pounds_\xi R_{\alpha \beta} 
\ee
where $e_{\mu \nu}$ are the coefficients of the Euler-Lagrange morphism given 
by equation (\ref{E1}).
From the  first-variation formula and the covariance condition it is possible 
to formulate the N\"other's theorem which associates to any vector field 
$\xi$ on the spacetime, a conserved current $\c{E} (L, \xi)$ so that:
\be
{Div}( \c{E} (L, \xi)) = \c{W} (L, \xi)  \label{cicci}
\ee
where $Div$ denotes the (formal) divergence operator.\\
The \textit{work-form} $\c{W} (L, \xi)$ vanishes on-shell, i.e. along 
solutions of field equations. For General 
Relativity in vacuum we obtain that:
\be
\c{E}^\lambda (L,\xi)=d_\mu  \Big[-  \frac{1}{2 k}   \sqrt{g} 
({\n_*}^\lambda \xi^\mu -{\n_*}^\mu \xi^\lambda) \Big] +(R_{\rho \nu}  - 
\frac{1}{2} \sqrt{g} g_{\rho \nu}) g^{\nu \lambda} \xi^\rho
\ee
For each natural theory (as well as for gauge-natural theories) using Bianchi's 
identities, it is possible to decompose the current $\c{E}$ as:

\be
\c{E} (L, \xi)= \widetilde{\c{E}} (L, \xi)+{Div}( {\c{U}} (L, \xi))
\ee
where $\widetilde{\c{E}}$ is defined the \textit{reduced current} and 
${\c{U}}$ is defined the \textit{superpotential} of the theory.
The reduced current $\widetilde{\c{E}}$ vanishes on-shell. For the 
Lagrangian (\ref{H}) the gravitational superpotential $\c{U}$ can be explicitely 
calculated (see \cite{FF1}) and it is known to be:
\be
\c{U} (L,\xi)=-  \frac{1}{2 k}   \sqrt{g} ({\n_*}^\lambda 
\xi^\mu -{\n_*}^\mu \xi^\lambda) ds_{\lambda \mu} \label{S1}
\ee 
which is called the \textit{Komar superpotential} \cite{FF1}. The  conserved quantities 
can now be obtained integrating the current $\c{E}$ on a $(n-1)$-region $D$, 
i.e. a compact submanifold of $M$ with a compact 
boundary $\pa D$. So the conserved quantities are integrals of 
the superpotential on $\pa D$. Generally the quantities obtained are not 
conserved with respect to the "time" defined by an ADM splitting of spacetime in 
spacelike surfaces. This 
happens, e.g., when the timelike vector $\xi$ is a Killing vector for 
the solution $g$ (see \cite{BY1}). We stress that all the above quantities 
are linear with 
respect to the vector $\xi$ (toghether with its derivatives).\\
However, if we calculate the conserved quantities for General 
Relativity  integrating the superpotential (\ref{S1}) on $\pa D$, the mass obtained does not assume the 
physically expected value. This is the well known \textit{anomalous  factor 
problem} which affects the Komar superpotential. There are at least two different 
ways to solve the problem \cite{FF1}. \\
If we consider the variation of conserved quantities, defined above, this 
expression will suggest us to define the variation of the corrected 
conserved quantities by the ADM prescription \cite{FF1}:
 \begin{eqnarray*}
\delta_X \widetilde{Q}_D (L,\xi,\sigma) =  \int_{\pa D}  [ \delta_X \c{U} (L,\xi, \sigma) 
- i_{\xi}   ( <  \F ( L, \gamma) \circ j^{2k-1} \sigma | j^{k-1} X >)]
\end{eqnarray*}
which gives us the expected quantities, of course up to an integration costant. 
To construct this formula in a covariant way it is 
necessary to introduce a  \textit{background connection} $\gamma$ and the conserved 
quantities will depend on the background connection chosen. We can consider 
$\gamma$ to provide us a "zero level" for the  energy, so that (as it is physically resonable) 
it is like a parameter for the theory. 
In the case of General Relativity we can choose as a sort of "natural" background connection 
the Levi-Civita connection of any \textit{background metric}.\\
On the other hand, one can notice that the Lagrangian (\ref{H}) of General 
Relativity  can be written as the  sum of a first order Lagrangian 
and a divergence depending on the background metric. The first order 
Lagrangian gives us a mechanism analogous to 
the ADM formalism (thought explicitely covariant and independent on the 
choice of a foliation) to calculate the corrected conserved quantities 
\cite{FF3}. The background fixing produces in this case an additional 
boundary term which solves the anomalus factor problem. The first order 
Lagrangian method is more general (in fact it is also applicable to non-compact 
solutions), while ADM formalism is not applicable in this case because space infinity 
is not a priori asymptotically flat. This is particularly important to our 
purposes, since cosmological solutions are usually non-compact and have 
general a different asymptotical structure.\\
\vskip3pt
Mass and angular momentum for a black hole solution of Einstein equations 
are defined as the conserved quantities  respectively connected to the 
vectors $\pa_t$ and $\pa_\phi$ on spacetime. The corrected quantities, 
which provide us the physically expected values,  are defined as integals at space infinity:
\begin{eqnarray}
{M}=\int_\infty [\c{U}_{Komar} (L,\pa_t, g) -B(L,\pa_t, g) 
\label{m+}\\
{J}=- \int_\infty [\c{U}_{Komar} (L,\pa_\phi, g)- B(L,\pa_\phi, g)]  \label{j+}
\end{eqnarray}
where the $(n-2)$-form $B$ is defined through integration of  the variational equation on $\pa D$:
\be
\delta_X B(L,\xi, g)=i_\xi {Div} <  \F ( L,\gamma) \circ j^3 
g | j^{1} X >
\ee
see for example \cite{FF2}, \cite{FF4}. \\
A covariant (and somehow canonical) choice for vacuum General Relativity is:
\be
 B(L,\xi, g)=- 
\sqrt{g} g^{\alpha \beta} \xi^{[\lambda} w^{\mu]}_{\alpha \beta} d s_{\lambda \mu}
\ee
where we have set $\gamma$ and $\Gamma$ to be the Chistoffel's symbols for 
the metric and the background connection respectively and we have defined:

$$
\cases{
w^{\mu}_{\alpha \beta} = u^{\mu}_{\alpha \beta}- U^{\mu}_{\alpha \beta} \cr
U^{\mu}_{\alpha \beta}=\Gamma^{\mu}_{\alpha \beta}-
\Gamma^{\rho}_{\rho (\beta} \delta^{\mu}_{\alpha)} \cr
u^{\mu}_{\alpha \beta}=\gamma^{\mu}_{\alpha \beta}-
\gamma^{\rho}_{\rho (\beta} \delta^{\mu}_{\alpha)}\cr
}
$$

\section{\large Definition of entropy in the stationary case}
The entropy of a stationary black hole solution is defined as the 
macroscopical quantity which satisfies the first principle of 
thermodynamics (\ref{prpr}). We impose that $X$ is a solution of linearized 
field equations and both the temperature $T$ and the angular velocity 
of black hole horizon $\Omega$ are constant parameters depending on the class of solutions chosen. The temperature $T$ is just defined as the temperature of the Hawking  
radiation $T=\frac{\kappa} {2 \pi}$, where $\kappa$ is the surface gravity, 
as shown in \cite{SH1}, \cite{BY1}, \cite{BY2}, \cite{GH1} by means of Euclidean path integrals.  On the other 
hand $\Omega$ is defined so that $\mid \xi \mid^2$ vanishes on the BH 
horizon. If we solve  (\ref{prpr}) with respect to $\delta_X \c{S}$ we obtain that:
\ba*
 \delta_X \c{S}=\frac{1}{T} (\delta M -\Omega \delta J) =
=\frac{1}{T}
  \int_{\infty} [\delta_X \c{U} (L,\xi, \sigma)-i_\xi   < \F  ( L, 
\gamma) | j^{1} X >] \label{ens}
\ea*
\\
where $\infty$ means the space infinity of a spacelike slice and 
$\xi=\pa_t+ \Omega \pa_\phi$. \\

Under the hypotheses that $\xi$ is a Killing vector for $g$ and $X$  
is  a solution of linearized field equations it is possible to prove 
in a very general framework (see \cite{FF2}) that the quantity under integral:
\be
\alpha(L,g,\xi,X)=\delta_X \c{U} (L,\xi, g)-i_\xi   < \F ( L, 
\gamma) | j^{1} X > \nonumber
\ee
is a closed form. This allows us to redefine $\delta_X \c{S}$ as an 
integral on any spatial surface which is homologically equivalent to 
$\infty$. In this definition we do not have any additional requirement about 
maximality of the solution considered neither about the horizon 
properties.  In particular it is not necessary to require $\xi$ to vanish on the
 trapping surface. 
This fact allows us to apply the definition to a wider range of solutions 
and simplifies both conceptually and computationally the calculations (see \cite{FF2}, 
\cite{FF9}, \cite{FF10}). \\
If the  solution admits a bifurcate Killing horizon and we can choose a 
bifurcation surface on which $\xi$ vanishes, then our more general definition reproduces, as 
a very particular case, the one given by Wald and Iyer in \cite{IW1}. 
This latter definition is 
not applicable to solutions for which $\xi$ is not a Killing vector (non-
stationary solutions, non-asymptotically flat solutions, etc...).

\section{\large Variation of conserved quantities}
It is possible to express a bundle morphism (for example the 
Euler-Lagrange and the Poincar\'e-Cartan morphisms which are $k$-forms on 
$M$) in local fibred 
coordinates. In this local formalism, any such morphism appears to be a linear 
combination of the vector field components $(\xi^\mu, \xi^i)$ together with their 
derivatives up to order $r$ ($r=1$ for the example of General Relativity 
under investigation). Let us thence consider a derivation $\delta$ (i.e. a linear operator which 
satisfies the Leibniz rule). If we are able to 
calculate the $\delta$-derivative of a vector field component and of the whole $n$-form 
then, 
by applying the Leibniz rule, we are able to define the derivative of the 
coefficients of the linear combination.\\
In our case we apply this rule to the Lie derivative and to the variation 
along the flow of a vector field $X$, which are both derivations.
For example, if we choose $\pounds_\xi$ as a particular derivation  onto the 
Poincar\'e-Cartan morphism, where $\xi$ is a vector field on spacetime and 
$X \in V(M)$, we obtain for a second order theory:
\be
\pounds_\xi <  \F ( L,\gamma) | j^{1} X >=\pounds_\xi \big[ 
{p}_i^{\lambda } {{X}^i}+{p}_i^{\lambda \mu} 
{X}^i_{\mu} \big] ds_\lambda \label{R4}
\ee
and applying the Leibniz rule we obtain:
$$
\cases{
\pounds_\xi {p}_i^{\nu}=
\big( d_\mu \xi^\mu {p}_i^{\nu}-d_\mu \xi^\nu 
{p}_i^{\mu}+\xi^\mu  d_\mu {p}_i^{\nu}+\pa_i \xi^j 
{p}_j^{\nu} \big) \cr
\pounds_\xi {p}_i^{\nu \rho}=
\big( d_\mu \xi^\mu {p}_i^{\nu \rho}-d_\mu \xi^\nu 
{p}_i^{\mu \rho}+
+\xi^\mu  d_\mu {p}_i^{\nu \rho}+\pa_i \xi^j 
{p}_j^{\nu \rho}-d_\mu \xi^\rho {p}_i^{\nu \mu} \big) \cr
}
$$
We remark that this definition for the Lie derivatives of the Poincar\'e-Cartan morphism 
can also be obtained by considering ${p}_i^{\nu},{p}_i^{\nu 
\rho}$ as the local expressions of a section on a suitable fiber bundle and 
thence applying the general definition (\ref{lie}) of Lie derivatives of sections of 
fiber bundles (see ref. \cite{KO1}, \cite{TR1}). \\
\vskip2pt
It is now possible to analyze the divergence of the form 
$\alpha(L,\sigma,\xi,X)$ in the case that no conditions whatsoever are imposed on the 
solution $\sigma$. In the case of stationary black holes one requires 
$\xi$ to be a Killing vector of the solution. Accordingly we assume in the 
general case that $\xi$ is a symmetry for $\sigma$, i.e. $\pounds_\xi \sigma=0$ 
which is  
fundamental to prove that $ {d} \alpha(L,\sigma,\xi,X)=0 $ (see \cite{FF2}). \\
If we relax this condition such a divergence does not vanish anylonger. 
This is the case of non-stationary black holes in a 
relativistic theory. In this general setting we have that:
\ba*
{Div} (\alpha (L, \xi, \sigma,X))&=&
{Div} (\delta_X \c{U} (L,\xi) -i_\xi   < \F  ( L, \gamma) | 
j^{k-1} X >)=\\
 &=&\delta_X < \F  ( L, \gamma)  | j^{k-1} \pounds_\xi \sigma > -\pounds_\xi  (<  \F ( L, \gamma ) \mid j^{k-1} X >)\\
&-&
\delta_X \widetilde{\c{E}} (L,\xi)-i_\xi 
<\E (L) | X>
\ea*
(see for example \cite{FF2}).\\
Let us thence analyze each term of this expression. The Euler-Lagrange 
morphism vanishes on-shell $<\E (L) | X>=0$. The variation of the 
reduced current can be expressed as:
\be
{Div} \delta_X \widetilde{\c{E}} (L,\xi)=\\
-\delta_X < \E (L) | \pounds_\xi \sigma>
\ee
and this term is identically vanishing since $X$ is a solution of the linearized 
field equation (see \cite{FF2}).\\
Using the prescription given for the variation of fiber bundle morphisms, 
in the case of a theory of order $k=2$, we can analyze the two terms left on 
the right hand side and we obtain as a special case:
\ba*
&{Div}& [\alpha (L,\xi,\sigma,X) ]=\omega(L,\xi,\sigma,X)=\\
 &=&<\delta_X  \F ( L ) | j^{1} \pounds_\xi \sigma>-
<\pounds_\xi \F ( L, \gamma ) | j^{1} X >+ < \F  ( L)  | j^{1} Z>
\ea*
where $\omega$ is an $(n-1)$-form on spacetime and $Z=Z^i \pa_i$ is a 
vertical field defined as 
$Z^i=\big( \pa_j X^i \pounds_\xi \sigma^j \big)$ so that it lifts to 
$j^1 Z=Z^i {\pa}_i+\big( d_\mu Z^i \big) {{\pa}_i}^\mu $. Let us stress 
that in the above expression each term is ``under 
control" in the sense that it can be analytically calculated whenever a 
Lagrangian is given for the theory. The expression of $ {Div} [\alpha 
(L,\xi,\sigma,X) ]$ is fundamental to our purpose; in fact it will 
contribute to the entropy formula under the form of a volume integral.\\
To summarize, in the case of under analysis (General Relativity in vacuum), 
we see that it is possible to calculate $\omega$ and $\alpha$, 
using the formula for the Komar superpotential and for the Poincar\'e-Cartan morphism, 
expressed in local coordinates by (\ref{S2}), (\ref{S1}), namely in our case:

$$
\cases{
\omega (L,\xi,\sigma,X)=\delta_X < \F ( L, \gamma) | j^{1} \pounds_\xi g>-\pounds_\xi < \F ( L, \gamma ) | j^{1} X >\cr
\alpha (L,\xi,\sigma,X)=\delta_X \c{U} (L,\xi, g)- i_\xi   < \F ( L, \gamma) \circ j^3 g| j^{1} X > 
}
$$
where variations and Lie derivatives can also be defined according to  (\ref{R4}) as usual 
for differential forms.

\section{\large Non-stationary black holes}
In this section we extend our definition for black hole 
entropy to the case of non-stationary black holes. As a motivation let us 
mention that cosmological solutions of Einstein's 
equations are usually not stationary and not asymptotically flat models, 
i.e. the solution does not admit any timelike Killing vector. In this case the 
$(n-2)$-form $\alpha$ is 
no longer closed and we cannot easily define the entropy as a boundary integral on a 
trapping surface for the singularity. \\
In our model the first principle of thermodynamics is the same used for 
stationary black 
holes. We consider black holes which do not emit gravitational waves, so we 
consider only solutions without a quadrupole momentum \cite{IW1}, 
\cite{FR1}, \cite{FR2}. 
This means that the 
system is isolated and electrically not charged. The geometrical formalism we use is manifestly
covariant. We will also show that our proposal satisfies all 
the reliability conditions stated by Wald and Iyer in \cite{IW1}.\\
We define again the entropy for a non-stationary solution of Einstein equations 
as the macroscopical quantity which satisfies the first principle of 
thermodynamics (\ref{prpr}). The definition is the same given before 
for the case of  stationary black holes. In this new case, however, $T$ and $\Omega$ 
cannot be calculated as the temperature and the angular velocity of black 
hole horizon, but they can be considered as a priori parameters of the 
theory. The only requirement is to ask these parameters to realize an 
integrable 
first principle of thermodynamics. The choice among them has to be carried 
over on the basis of some external physical consideration. 
 However this is not a feature of non-stationary solutions;
 even in the stationary case, if we choose quasi-local energy instead 
of mass we obtain 
a different (but integrable) first principle \cite{BY1}. This fact 
will be subject of further investigations.\\
On the other hand, the mass and the angular momentum may no longer be 
time-conserved on a spacelike ADM slice of spacetime, but they are covariantly 
conserved, i.e. they obey a continuity equation; in other words they are 
conserved in the sense of N\"other theorem even if their 
values may change in time.\\
We can substitute in the first principle the expressions (\ref{m+}) and 
(\ref{j+}) for mass and angular momentum calculated by means of N\"other theorem 
and we will obtain an expression 
which defines the variation of entropy as an integral on space infinity (\ref{ens}). Now, 
in the case  of non-stationary black holes, we have to take into account 
that $\alpha$ is not closed to evaluate the same quantity on a trapping 
surface. We have to consider the form $\omega$ and its integral on a 
volume $\Sigma$ \textit{between} the trapping surface $\Pi$ and the space infinity. 
So we will obtain the formula for entropy under the form:
\be
T \delta_X S_{dyn}= \int_\Pi \alpha(L,g,\xi)+ \int_\Sigma 
\omega(L,g,\xi) \label{U2}
\ee
where $T$ is the black hole temperature and the 
integrated forms $\omega $ and $ \alpha$ are defined in the previous section. In this 
formula each term is explicitly known and computable once the Lagrangian 
and the exact solution $g$ are specified. The only restriction 
on the theories we analyze is the fact that they have to be well-defined 
from a Lagrangian viewpoint.\\
The effort to apply this formula to known solutions has been vain up to now.
In the case of General Relativity in vacuum, in fact, to our knowledge there are no well defined exact 
non-stationary solutions in literature, even if the Hilbert Lagrangian which defines 
the theory is 
 global and covariant. If we consider otherwise the case of General Relativity 
in interaction with matter it is then possible to find in literature some explicit 
non-stationary solution; in this case 
it would be easily possible to generalize the definition of entropy (\ref{U2}) to treat also 
these theories by just
 adding an interaction term to the superpotential which eventually enters the final formula (\ref{U2}) (see \cite{IW4}). 
In this latter case, however, 
there is no well defined global Lagrangian for the theory,
 because of the exhotic properties of the gas matter considered, which is an essential 
requirement to calculate conserved quantities in a geometrical framework. 
However we carry over a theoretical analysis of the case of General Realtivity in vacuum.
\vskip2pt
Wald and Iyer imposed some conditions \textit{a priori} on the reliability 
of the definition which our 
definition satisfies by default:\\
$\bullet$ In the case of stationary black holes we must have $T \delta_X \c{S}=T 
\delta_X \c{S}_{dyn}$. To show this fact it is enough to say that when the 
solution is stationary we have
 $\pounds_\xi g=0 \Rightarrow \omega(L,g,\xi)=0$. \\
$\bullet$ We have to show that in the case of non-stationary perturbations, 
generated by a field $\widetilde{X}$, of a stationary solution we have 
$T \delta_{ \widetilde{X}} {S_{dyn}}=T \delta_{ \widetilde{X}}{S}$. In 
this case once again $\omega(L,g,\xi)=0$. This is related to the fact 
that what we need is $\pounds_\xi g_0 =0$ where $g_0$ is the unperturbed 
solution, which is stationary.  \\
$\bullet$ The entropy for a theory  defined by an equivalent Lagrangian 
$L+{Div} \theta$ 
should be the same calculated for the theory defined by the Lagrangian 
$L$. It easy to see that for any pure divergence Lagrangian we have
 $\alpha( {Div}\theta,\sigma,\xi)=0$. Since $\alpha$ 
is linear in $L$, in fact we have that $\alpha=0 
\Rightarrow \omega=0$ because we have chosen a pure divergence Lagrangian. \\
$\bullet$ Our definition must satisfy the second principle. This point is 
left out to future investigations, but we stress that at the moment the problem is still 
out of control even in 
the stationary case
\cite{IW1}; however it is reasonable to say that the second principle is 
related to 
the second variation of the fibered morphisms we have constructed, and thence 
to the positivity of energy.\\
$\bullet$ Finally, our definition should be covariant under field redefinition. Our 
formalism is manifestly covariant by construction and our definition 
satisfies this condition too. It is easy to prove this claim if we 
consider the trasformation rules for the Poincar\'e-Cartan morphism 
coefficients and for the field $j^{k-1} X$.\\

\section{\large Conclusions and perspectives}
We have proposed a new prescription to calculate the entropy for a non-stationary 
black hole. This formula is applicable to a well defined relativistic 
theory with a known (global) Lagrangian and whenever Lie derivatives are 
well-defined so that we can define covariant conserved quantities at space infinity. 
The application of our formalism 
to calculate entropy for a well-defined explicit solution is immediate; once we have 
the solution in a local coordinate system it is possible to apply the algorithmical 
formalism we have developed to calculate the conserved quantities and entropy. 
The formula we have proposed is independent on the choice of 
the trapping horizon for the singularity; if we consider horizons belonging to the same 
homotopy class, 
the result obtained for entropy is invariant (see for example the TAUB-BOLT solution,
 a discussion of which is given in \cite{FF10}). \\
From a physical viewpoint this definition satisfies the conditions 
stated by Wald and Iyer in \cite{IW1} and in particular it is covariant. A future 
task will be to calculate explicitly the entropy for some exact non-stationary solution 
of Einstein equations. We will furthermore investigate the second principle 
for our definition both for stationary and non-stationary black holes. 

\section{\large Acknowledgements}
We are grateful to M. Ferraris and M. Raiteri for fruitful discussions 
and for their helpful suggestions.

\end{document}